\documentclass[aps,prl,twocolumn]{revtex4-1}

\usepackage{graphicx}
\usepackage{amssymb}
\usepackage{amsmath}
\usepackage{cleveref}
\usepackage{bbold}
\usepackage{epstopdf}
\usepackage{verbatim}
\usepackage{nicefrac}

\crefname{equation}{Eq.}{Eqs.}
\crefname{figure}{Fig.}{Figs.}

\newcommand{\bms}[1]{\boldsymbol{#1}}

\newcommand{\vphi}{\varphi}
\newcommand{\bn}{\boldsymbol{n}}
\newcommand{\ca}{{\rm Ca}}
\newcommand{\ch}{{\rm C_h}}
\newcommand{\tv}{\boldsymbol{v}}
\newcommand{\por}{p}
\newcommand{\res}{r}
\newcommand{\ur}{u_\res}
\newcommand{\up}{u_\por}
\newcommand{\rp}{R_\por}
\newcommand{\rr}{R_\res}
\newcommand{\lp}{L_\por}
\newcommand{\lr}{L}
\newcommand{\pep}{{\rm Pe}_\por}
\newcommand{\per}{{\rm Pe}_\res}
\newcommand{\tnabla}{\nabla}
\newcommand{\vpavg}{\left<\vphi\right>}
\newcommand{\vpout}{\vpavg_{{\rm out}}}

\newcommand{\jd}{\boldsymbol{j_d}}
\newcommand{\ja}{\boldsymbol{j_a}}
\newcommand{\muex}{\mu_{{\rm eff}}}

\begin{document}

\title{Interplay between adsorption and hydrodynamics in nanochannels: 
towards tunable membranes}

\author{Sela Samin}
\email{S.Samin@uu.nl}
\author{Ren\'{e} van Roij}

\affiliation{Institute for Theoretical Physics, 
Center for Extreme Matter and Emergent Phenomena, 
Utrecht University, Princetonplein 5, 3584 CC Utrecht, The Netherlands}

\date{\today}


\begin{abstract}
We study how adsorption of a near-critical binary mixture in a nanopore is 
modified 
by flow inside the pore. We identify three 
types of steady states upon variation of the pore P\'{e}clet number ($\pep$), 
which 
can be reversibly accessed by the application of an external 
pressure. Interestingly, for small $\pep$ the pore acts as a weakly selective 
membrane which separates the mixture. For intermediate $\pep$, the 
flow effectively shifts the adsorption in the pore thereby opening 
possibilities for enhanced and tunable solute transport through the pore. For 
large $\pep$, the adsorption is progressively reduced inside the 
pore, accompanied by a long ranged dispersion of the mixture far from the pore.

\end{abstract}
\maketitle 

Pressure-driven membrane processes are widely used in water treatment 
technology 
\cite{pendergast2011}. With the aim of 
producing more energy-efficient and eco-friendly membranes, intense ongoing 
research is dedicated to the improvement of the membrane's conductivity and 
selectivity \cite{powell2011,carta2013,wang2015} and to prevent fouling 
\cite{hou2015}. Yet, advances are impeded by the lacking understanding of 
transport in membranes used for removing various organic contaminants or oil 
spills from water \cite{adebajo2003}, in pressure-driven hydrocarbon 
recovery in nanoporous rocks \cite{monteiro2012,lee2016}, and organic solvent 
nanofiltration \cite{marchetti2014,amirilargani2016,gravelle2016}.

Also from a fundamental perspective, the role of hydrodynamics in phase 
separation kinetics \cite{onuki1997} near 
wetting surfaces \cite{tanaka2001} and in confinement 
\cite{Binder2010,bocquet2010} has 
been extensively studied, as well as its role in wetting of immiscible 
oil-water systems in micro- and nano-channels \cite{rauscher2008}, and in 
nanobubbles formation \cite{lohse2015}. More recently, capillary driven flows 
in nanometric channels have been extensively studied 
\cite{huber2015,vincent2016}, also in mixtures \cite{oh2009}. 

The consequences of pressure-induced flow across a nano-channel 
and perpendicular to a wetting or adsorption 
layer have been studied before in the context of electrolyte 
\cite{heyden2005,heyden2006,lis2014} and polymer \cite{khare2006} solutions. 
However, the consequences for a non-ideal liquid mixture, where the 
hydrodynamics-adsorption coupling of the solvent itself is 
temperature- and composition-sensitive, has not been explored before to the 
best of our knowledge. This scenario 
is, however, common in \emph{any} membrane process involving liquid mixtures and 
also bears resemblance to the classical Graetz problem of mass (or heat) 
transfer into a fluid in steady flow through a cylindrical tube 
\cite{graetz1882,deen2011}. Nevertheless, there are 
important differences. Unlike the Graetz problem, there 
is no net flux of material from a wetting wall into the fluid. More 
importantly, when the adsorbed fluid is the solvent itself, in contrast to the 
dilute gas typically considered in the Graetz problem and general mass transfer 
problems, an important length scale appears, namely, the correlation length 
$\xi$. In this Letter, we show that when $\xi$ is of the order of the channel 
size, hydrodynamics and wetting become strongly coupled, leading to several 
surprising effects. We focus on the critical adsorption regime 
\cite{fisher1978}, in the one-phase regime close to the critical temperature 
$T_c$ of the mixture, where $\xi$ is naturally large. However, our results 
should be applicable whenever the thickness of the adsorbed film is 
comparable to the channel size, also far from $T_c$.

\begin{figure}[!tb]
\includegraphics[width=2.5in,clip]{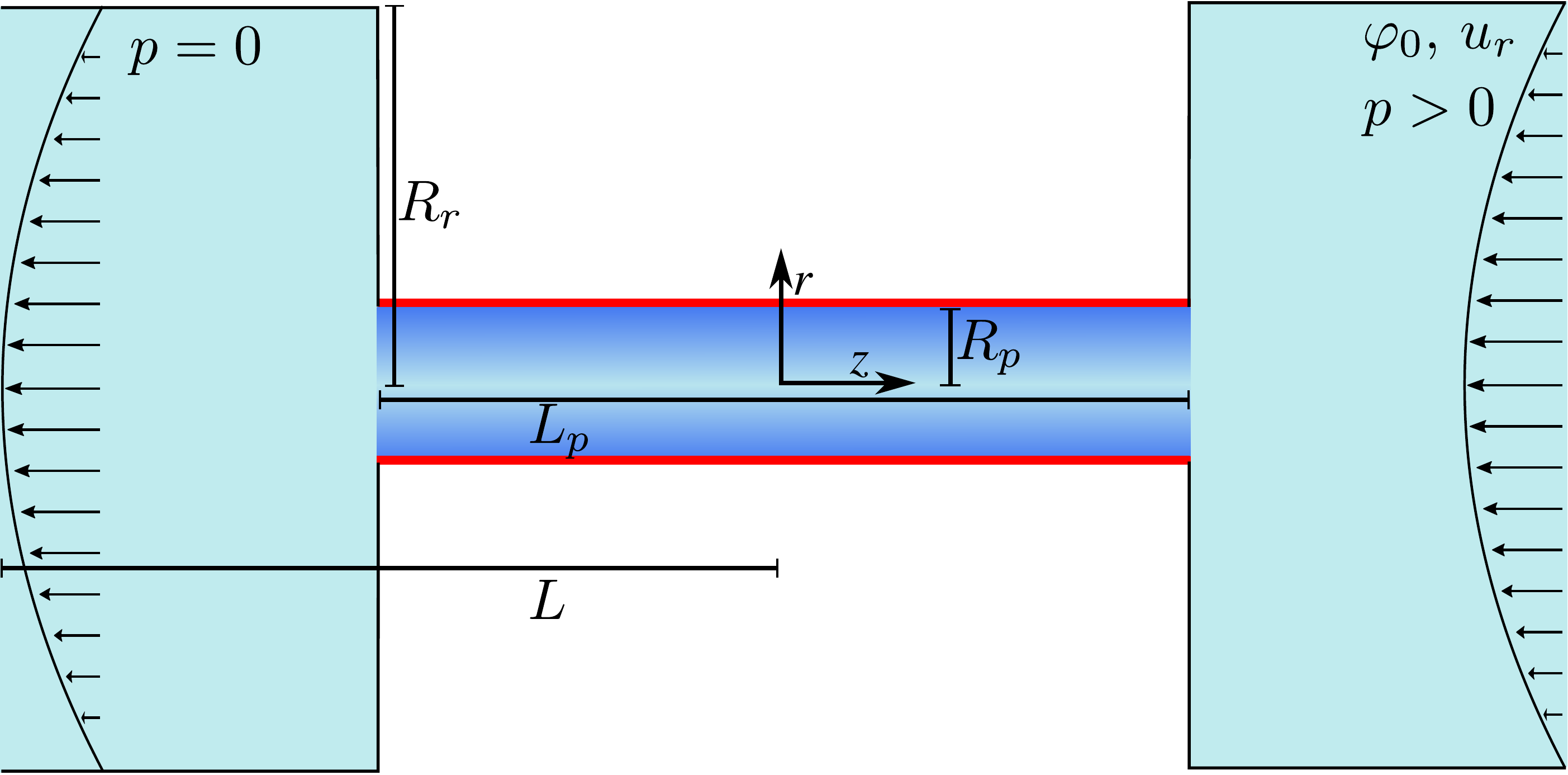}
\caption{Schematic illustration of a cylindrical pore with radius $\rp$ and 
length $\lp$ connecting two reservoirs with radius $\rr$ and 
length $\lr-\lp/2$. In equilibrium, the reservoirs' compositions are $\vphi_0$ 
but 
inside 
the pore the composition is larger due to solvent adsorption to the attractive 
pore wall (red line). After the application of pressure to the right 
reservoir, a Hagen--Poiseuille flow with a parabolic velocity profile and an 
average velocity $\ur$ develops far from the 
pore.}
\label{fig_sys}
\end{figure}

Using Direct Numerical Simulations 
(DNS) we explore the effect of solvent adsorption to channel walls in 
nanofluidic systems 
of binary mixtures. We consider a near-critical binary mixture characterized by 
the order parameter $\vphi$, which denotes the deviation of the mixture 
volume fraction from 
its critical value $\vphi_c$. By near-critical we mean that the mixture 
temperature $T$ 
and average composition $\vphi_0$ are in the neighborhood of the 
critical point $(T_c,\vphi_c)$. Two 
cylindrical material 
reservoirs with radius $\rr$ and length $\lr-\lp/2$ containing the mixture are 
connected by a narrow cylindrical pore with radius $\rp$ and length $\lp$, see 
\cref{fig_sys}. In equilibrium, the composition 
of both reservoirs is $\vphi_0$ but inside the pore $\vphi>\vphi_0$ is not 
uniform since the pore wall favors one component. Close enough to $T_c$ the 
mixture 
correlation length $\xi$ becomes comparable to $\rp$ and $\vphi>\vphi_0$ 
throughout the pore volume \cite{marconi1988,maciolek2003} due to adsorption. 
We then impose a pressure at the edge of right reservoir in 
\cref{fig_sys}, forcing the mixture through the pore and leading to a 
fully-developed Hagen--Poiseuille flow with a mean velocity $\ur$ far from the 
pore.  
For an incompressible fluid, the mean velocity in the pore is 
hence $\up=(\rr/\rp)^2\ur$.

To investigate the dynamics and steady-state we employ a classical 
continuum framework \cite{anderson1998}. The time evolution of the fluid is 
given by the so-called model-H equations, 
combining the convective Cahn-Hilliard equation for the 
composition with the Stokes equations for the fluid 
velocity for small Reynolds number, ${\rm 
Re}\ll1$. In 
dimensionless form the governing equations read 
\begin{align}
\label{eq:ch_dl}
 \partial\vphi / \partial t &=- \tnabla \cdot \left(\pep \vphi 
\tv-\tnabla \mu\right)~,\\
\label{eq:mu_dl}
\mu&= -\epsilon \nabla^2\vphi+f'(\vphi)~,\\
\label{eq:nc_dl}
\tnabla \cdot \tv &=0~,
\\
\label{eq:ns_dl}
\tnabla\cdot \boldsymbol{\tau} &= 
\nabla p+ \ch^{-1}\ca^{-1}\vphi \tnabla \mu~.
\end{align}
Here, all lengths are scaled by $\rp$, the velocity $\tv$ by $\up$, the 
chemical potential $\mu$ by the thermal energy $k_BT$, and 
time is scaled by $\rp^2/D$, where $D$ is the mixture inter-diffusion 
constant. Three dimensionless groups appear in this form of the equations. The 
most important is the P\'{e}clet number in the pore, $\pep=\up 
\rp/D$, measuring the relative magnitude of the composition advective current, 
$\ja=\pep \vphi 
\tv$, and diffusive current, $\jd=-\tnabla \mu$, in \cref{eq:ch_dl}. The 
chemical 
potential is derived from a Ginzburg-Landau free-energy for a LCST-type 
mixture \cite{com_sup}, where 
the Laplacian term in \cref{eq:mu_dl} accounts for composition 
inhomogeneities and the bulk part is given by 
the derivative of $f=\alpha\vphi^2/2+4\vphi^4/3$, where $\alpha=2(\chi-2)$ with
$\chi \sim 1/T$ the 
Flory 
interaction parameter. Composition gradients are characterized by 
$\epsilon=\chi\ch^2$ \cite{safran}, where $\ch=a/\rp$ is the Cahn number, with 
$a$ a 
molecular length characterizing both mixture components. In the 
Stokes 
equations, \cref{eq:ns_dl,eq:nc_dl}, $p$ and 
$\boldsymbol{\tau}=\nabla\bms{v}+\nabla\bms{v}^T$ 
are the dimensionless fluid pressure and viscous stress tensor, respectively, 
scaled by $\eta\up/\rp$, where $\eta$ is the 
fluid viscosity. The last term in 
\eqref{eq:ns_dl} is a body force due
to chemical potential gradients, which is inversely proportional to the  
capillary number, $\ca=a^2\eta\up/k_BT$, measuring the 
relative magnitude of viscous and interfacial forces.

We use a cylindrical system of coordinates $(r,z)$ with 
$z\in[-\lr,\lr]$ 
and $r\in[0,\rr]$, and employ symmetry boundary 
conditions (BCs) at $r=0$. At the inlet ($z=\lr$) we impose a critical 
composition with $\vphi_0=\vphi_c=0$, the 
corresponding chemical potential $\mu_0=0$ and a fully-developed laminar 
flow with a mean velocity $\ur$ (see \cref{fig_sys}). At the outlet ($z=-\lr$), 
we allow the mixture 
to be freely advected \cite{com_sup}. 
On all other solid boundaries we impose 
no-slip for the velocity $\tv=0$ \cite{com_slip} and no composition flux, $\bn 
\cdot  \nabla\mu=0$, 
where $\bn$ is the outward unit vector normal to the surface. 
At the reservoir walls $\bn \cdot \nabla\vphi=0$, as there is no adsorption of 
either component. The 
pore wall, however, does preferentially adsorb one of the components, and 
therefore $\bn \cdot \nabla\vphi|_{r=\rp}=\gamma/\epsilon$ for 
$\left|z\right|<\lp/2$, where $\gamma$ 
measures 
the difference 
between the short-range interaction of the two solvent components 
and the solid \cite{com_bc}. We use 
$\gamma=0.1\ch$ that leads to weak 
critical adsorption \cite{cho2001}.

For concreteness, we use the physical properties of the experimentally common 
mixture  water--2,6-lutidine \cite{com_sup,grattoni1993}. Using 
typical values leads to $\ca\sim10^{-6}-10^{-4}$ and $\ch\sim0.1-0.01$ for 
pore mean velocities $\up\sim 0.1-10$ mm/s and pore 
radii $\rp\sim 1-10$ nm. Unlike other properties, the inter-diffusion 
constant is sensitive to temperature, and close to $T_c$ it follows 
the Stokes-Einstein-Kawasaki-Ferrell relation 
\cite{kawasaki1970} $D=k_BT/(6\pi\eta\xi)$, where the correlation 
length $\xi$ in our mean-field description follows the scaling 
$\xi\propto(\left|T-T_c\right|/T)^{-\nicefrac{1}{2}}$ \cite{com_deb}.
Hence, for $\up$ 
mentioned above and for $T-T_c\sim 1-10$ K we find 
$\pep\sim0.01-1$, that is, diffusion and advection contributions are 
comparable. In the 
reservoirs $\per$ is smaller by a factor $\rr/\rp\sim10-100$ and hence 
transport in the reservoirs can be diffusion dominated.

\begin{figure}[!tb]
\includegraphics[width=3.5in,clip]{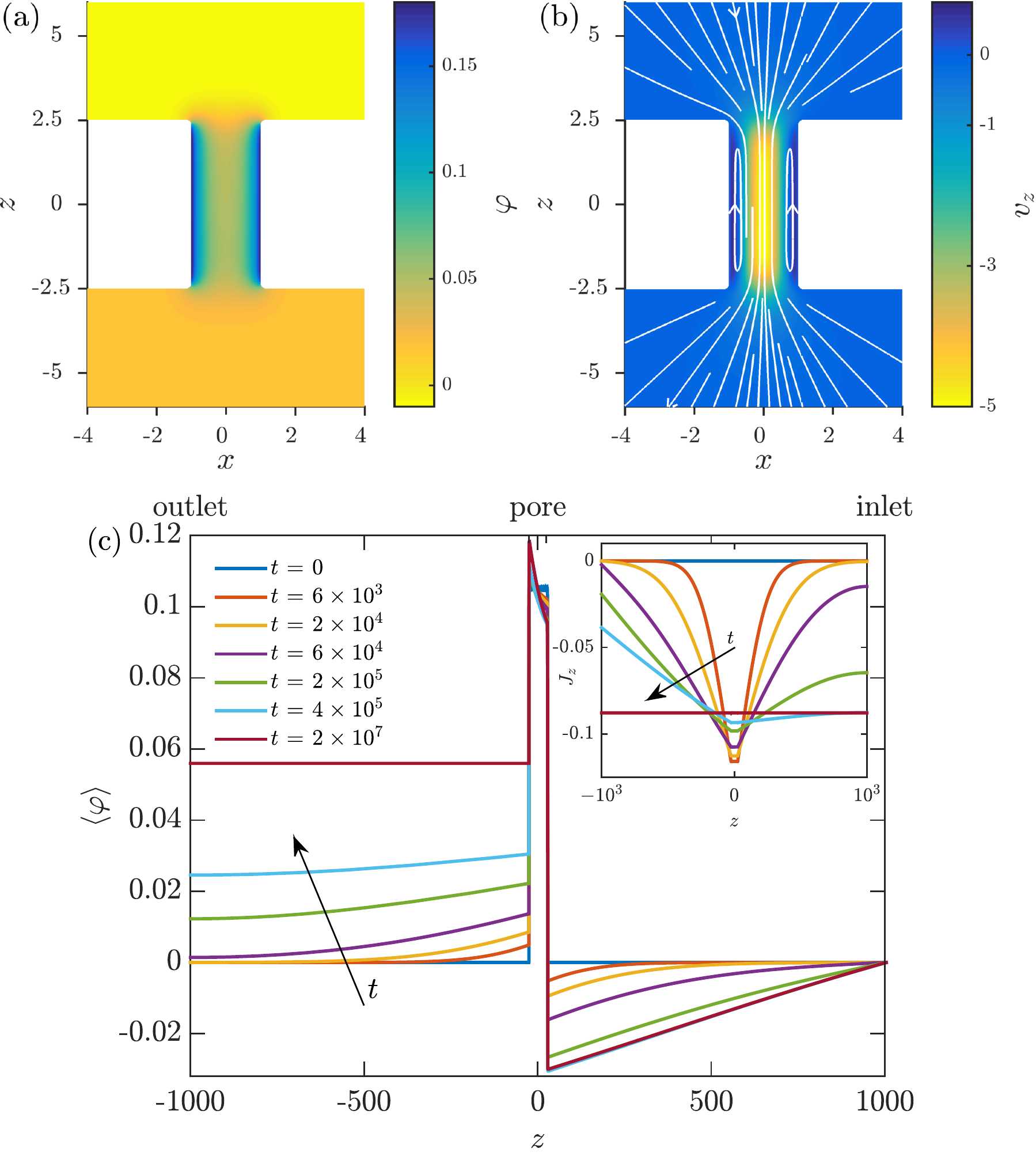}
\caption{(color online) (a) composition and (b) 
velocity $z$-component and streamlines for steady state case (i) in the $xOz$ 
plane for a 
mixture at a 
temperature $T_c-3$ K ($\xi=4.86$ nm) and with $\up=0.05$ mm/s ($\pep=0.0077$). 
The unit of length 
in all figures is 
the pore radius $\rp=5$ nm, and here we also set $\lp=5$, $\rr=20$ and 
$\lr=10^3$, i.e., the reservoirs' edges and their side walls are (far) 
beyond the scale of this plot. (c) Time evolution of the radially averaged 
composition $\vpavg$ profiles for case (i) using same the parameters as in (a) 
but with $\lp=50$. Inset: the corresponding $z$-component of the composition 
current $\bms{J}_z$ (see text).}
\label{fig2}
\end{figure}

DNS of the model-H equations \cref{eq:ch_dl,eq:mu_dl,eq:nc_dl,eq:ns_dl} in 
the bulk one-phase region up to 20 K from $T_c$, and with $\rp=5$ nm, reveals 
three representative 
steady states with increasing $\pep$: (i) at small 
$\pep$, the adsorption in the pore, $\Gamma=\int_{pore}[\vphi-\vphi_0]{\rm 
d}^3\bms{{\rm r}}$, 
remains close to its equilibrium value, (ii) at an intermediate $\pep$ range, 
the adsorption in the pore is reduced to almost constant as a function of 
$\pep$, and (iii) at large
$\pep$, the adsorption in the pore progressively decreases with $\pep$. 
In this Letter we focus on the first two cases. 

In case (i), which is realized at $\pep\lesssim0.1$, the composition 
distribution in the pore changes very 
little compared to the classical 
equilibrium case \cite{com_sup}, with 
$\vphi>\vphi_c$ throughout the pore as can be seen in \cref{fig2}(a). Most 
strikingly however, \cref{fig2}(a) also shows that the composition 
\emph{outside} 
the pore is \emph{not} equal to $\vphi_c$ but is uniform and higher than 
$\vphi_c$ in 
the outlet reservoir and non-uniform and smaller than $\vphi_c$ in the inlet 
reservoir. 
This implies that there is a steady current of composition through the 
pore, 
which effectively acts as a selective membrane as it lets only a higher 
composition 
mixture to leave. To understand the origin of this intriguing effect we plot in 
\cref{fig2}(c) the time evolution of the profiles of the radially averaged 
composition, $\vpavg(z)=2\pi\int\vphi(r,z) r {\rm d} r/A(z)$, where $A(z)$ is 
the
cross section area. For visual clarity, we plot in \cref{fig2}(c) results for a 
longer 
pore with $\lp=50$ but otherwise all the parameters are the same as in 
\cref{fig2}(a). \cref{fig2}(c) shows that shortly after the application of the 
pressure, a small excess composition is advected from the pore to the outlet 
reservoir; this excess composition diffuses quickly in the outlet, where 
the P\'{e}clet number is much smaller. The resulting depletion of composition 
in the pore is compensated by a predominantly diffusive current from the inlet 
reservoir, leading in turn to the small depletion in composition near the pore 
inlet. As time progresses, more of the excess composition in the pore is 
advected downstream and the outlet composition steadily increases, while at the 
same time the inlet reservoir is depleted. The inset of \cref{fig2}(c) shows 
the 
corresponding $z$-component of the composition current, 
$\bms{J}_z=2\pi\int \left[\ja_{,z}+\jd_{,z} \right] r {\rm d} 
r$, revealing that the composition current from the inlet grows 
until it is able to compensate the current out of the pore. When this occurs 
the excess composition in the outlet eventually saturates at a constant steady 
state composition $\vpout$. The value of $\vpout\approx0.06$ in 
\cref{fig2}(c) is a significant deviation from the critical composition, 
especially considering that in our weak adsorption model the 
average excess composition in the pore is only $\approx0.1$.

%
%

In short, adsorption in pores at small 
$\pep$ allows to separate out the binary
mixture component that is preferably adsorbed in the pore. Thus, 
relatively large nanopores can 
effectively act as a weakly selective membrane, by applying only a 
small pressure of $p<1$ bar, with possibly an extremely low energy 
consumption in membrane processes \cite{marchetti2014}. What enables this 
extraordinary behaviour is the 
interplay 
between advection and diffusion that develops \emph{inside} the pore at 
$\pep\lesssim0.1$. In 
this $\pep$ regime, a diffusive flux, counteracting partially the advective 
flux, is able to develop in order to restore the energetically favorable 
equilibrium adsorption. However, at finite $\pep$ there is always a residual 
advective flux and an excess composition in the outlet at 
steady state as a byproduct. The existence of a steady-state relies on the 
inlet 
reservoir 
being able to supply the constant current of excess composition, which in our 
model is guaranteed by constraining $\vphi=0$ at $z=\lr$, with this constraint 
satisfied naturally for $\lr\rightarrow\infty$ or at least for a 
significant duration of time even when the reservoir can be 
depleted \cite{com_sup}.

At steady state, the diffusive current inside the pore is almost uniform 
(except near the pore mouths), and points in the 
$\hat{z}$ direction, opposite to the incoming flow \cite{com_sup}. This leads 
to 
a uniform body force in the same direction which is most significant near the 
pore walls where the local capillary number is small. This creates a 
back-flow near the wall, resulting in a far-from--parabolic velocity 
profile inside the pore. Since the volumetric flow rate is 
constant this implies two planes of zero velocity and an increase of the flow 
in the center. These features are visualized 
in \cref{fig2}(b), which shows $v_z$ and streamlines corresponding to 
\cref{fig2}(a), and in \cref{fig5}(b), which shows the large deviation from  
Hagen--Poiseuille flow.
\begin{figure}[!tb]
\includegraphics[width=3.5in,clip]{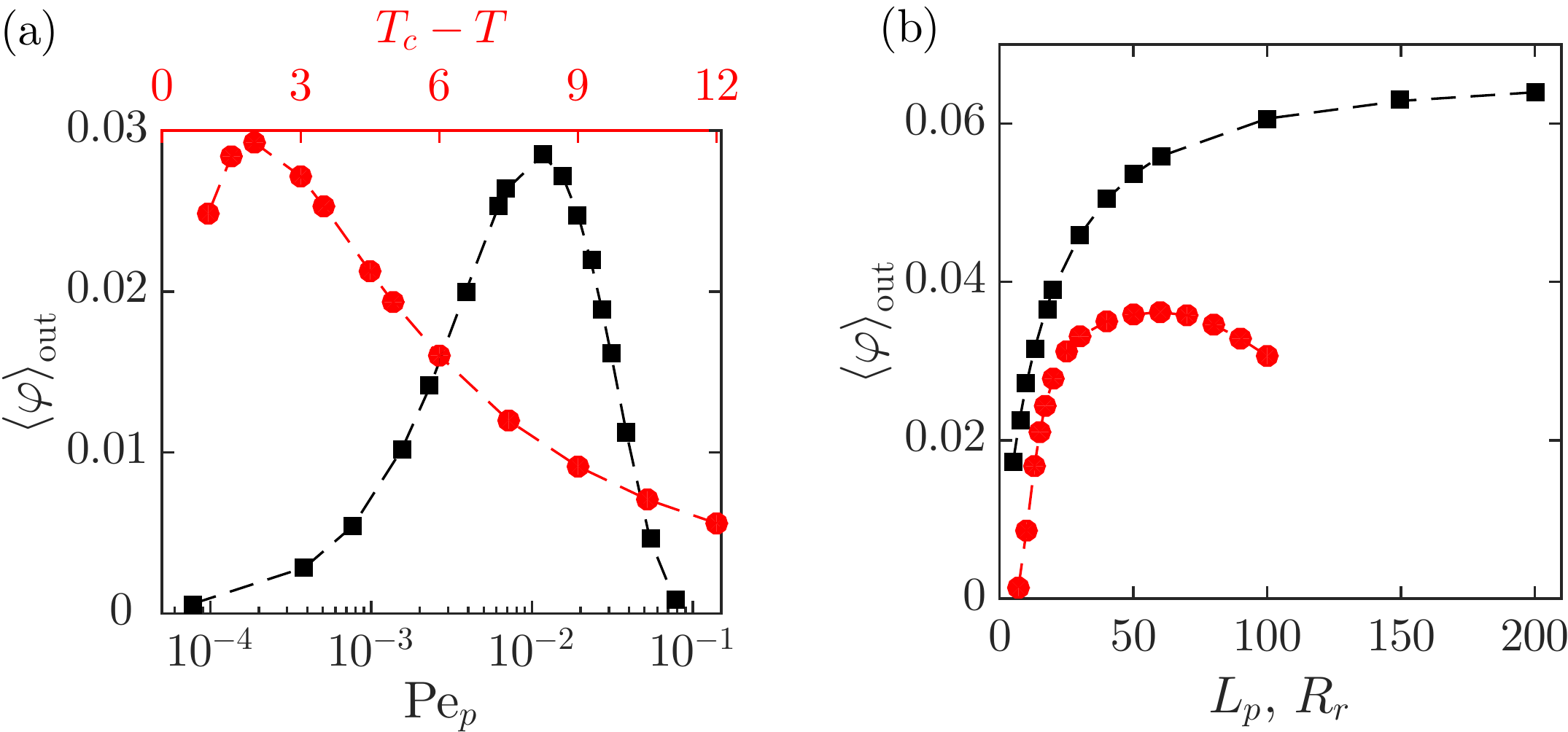}
\caption{(a) dependence of the outlet reservoir composition 
$\vpout$ on the pore P\'{e}clet number (squares, black abscissa) and deviation 
from the critical temperature (circles, red abscissa). (b) dependence 
of $\vpout$ 
on the pore volume and inlet reservoir volume. The pore 
volume changes linearly by varying
$\lp$ (squares) and keeping $\rp$ fixed and the reservoir volume changes 
quadratically by varying $\rr$ (circles) and keeping $\lr$ fixed. 
In both panels, the basis parameter set is $T=T_c-3$ K, 
$\up=0.05$ mm/s ($\pep=0.0077$), $\lp=10$, $\rr=20$ and $\lr=10^3$, excluding 
the 
abscissa parameter. Lines are a guide to the eye.}
\label{fig4}
\end{figure}

In \cref{fig4} we show the effects of varying some of the free parameters of 
our system on the outlet reservoir composition $\vpout$. \cref{fig4}(a) 
verifies the small $\pep$ regime (squares, black abscissa), showing that 
$\vpout$ increases with 
$\pep$ as more material is advected out of the pore but then vanishes gradually
beyond $\pep\sim0.1$ when axial diffusion can no longer maintain a steady 
current. 
\cref{fig4}(a) also shows that a temperature that maximizes $\vpout$ exists 
(circles, red abscissa) because close to $T_c$ $\pep\propto D^{-1}$ becomes 
very 
large 
whereas far from $T_c$ the adsorption in the pore is diminished. Note that 
$\vpout$ decreases slowly for large 
$T_c-T$ and is still significant up to 10 K from $T_c$. The effect of 
increasing $\lp$ on $\vpout$ is plotted in \cref{fig4}(b) with $\rp$ fixed 
(squares). For small $\lp$, $\vpout$ 
increases 
rapidly but then saturates when transport within the pore 
becomes the limiting factor, showing that above $\lp\sim100$ gains in $\vpout$ 
are marginal. Changing 
$\rr$ at fixed $\lr$ (\cref{fig4}(b), circles), we find that an 
optimal value of $\rr$ exists. Since we fix also $\pep$ here, the reservoir
P\'{e}clet number $\per$ decreases linearly with $\rr$. In the limit of a 
single long channel ($\rr\rightarrow1$), $\per\rightarrow\pep$ and advection in 
the outlet reservoir is dominant, leading to $\vpout\rightarrow0$. When $\rr$ 
becomes very large, $\per\rightarrow0$, reducing the total current from the 
inlet which leads to a decrease in $\vpout$. The results in \cref{fig4} 
should serve a 
guideline for the future design and optimization of membranes. We speculate 
that similar results would be obtained whenever 
a wetting film comparable in size to $\rp$ exists, for example, when 
capillary condensation occurs.

Upon increasing the external pressure the near-equilibrium 
adsorption cannot be maintained in any cross section along the pore axis, and a 
new type of steady state, case (ii), is reached. In \cref{fig5}(a) we show a 
representative 
composition map for $\pep=0.77$, where we used 
the same parameters as in 
\cref{fig2}(a) but increased $\ur$. Here, the 
adsorption at the pore wall is still significant 
and not far from the equilibrium value. However, $\vphi$ decays in the radial 
direction to a value that is smaller than $\vphi_c$ at the 
pore center (which is not possible in equilibrium within our continuum 
description). In this steady state the critical composition flows 
uninterrupted from the inlet reservoir to the outlet as there is no net 
composition current through the system. 
Furthermore, our DNS show that both the axial diffusive and axial advective 
current in the pore vanish. Having $\jd_{,z}=0$ implies a uniform chemical 
potential in the pore and therefore no body-force on the fluid. Thus, the 
velocity profile has a simple parabolic shape (see 
\cref{fig5}(b)) and being unidirectional, 
$\ja_{,z}=0$ means there must be a region of $\vphi<\vphi_c$ 
for some $r$. Hence, the flow effectively shifts the chemical potential to a 
smaller value $\muex<0$, which, since $\muex$ is constant, corresponds to an 
equilibrium adsorption for some $\vphi_0<\vphi_c$. 

An estimate of $\muex$ can be
calculated within a standard approximation neglecting the 
$\vphi^3$ term in $\mu$ \cite{onuki_book}. Ignoring small axial 
components inside the pore the composition profile is obtained from 
a simplified 
\cref{eq:mu_dl} which reads $\muex= 
-\epsilon(\vphi_{rr}+\vphi_r/r)+\alpha\vphi$.
This equation is solved together with the BCs $\vphi_r(0)=0$, 
$\vphi_r(1)=\gamma/\epsilon$ and the constraint  
$\ja_{,z}=2\pi\int_0^1
\left[-2\pep(1-r^2)\vphi(r)\right] r {\rm d}r=0$, which determines 
$\muex$. The resulting composition profile is
\begin{align}
 \vphi(r)&=\gamma/(\epsilon\zeta I_1(\zeta))\times 
\left[I_0(r\zeta)-8I_2(\zeta)/\zeta^2\right]~,
\label{eq:prof}
\end{align}
where $\zeta=\sqrt{\alpha/\epsilon}=\sqrt{2}\rp/\xi$ and $I_n$ are $n$-th order 
modified Bessel functions of the first 
kind. The first term in the brackets is the adsorption 
profile with no flow 
while the second term is the constant $r$-independent negative shift of $\vphi$ 
due to the 
flow. Strikingly, this result is independent of $\pep$. Indeed, our 
DNS confirm that the composition profile in the pore is unchanged for the 
parameters of \cref{fig2}, in the range $0.5 \lesssim 
\pep \lesssim 10$.

\begin{figure}[!tb]
\includegraphics[width=3.5in,clip]{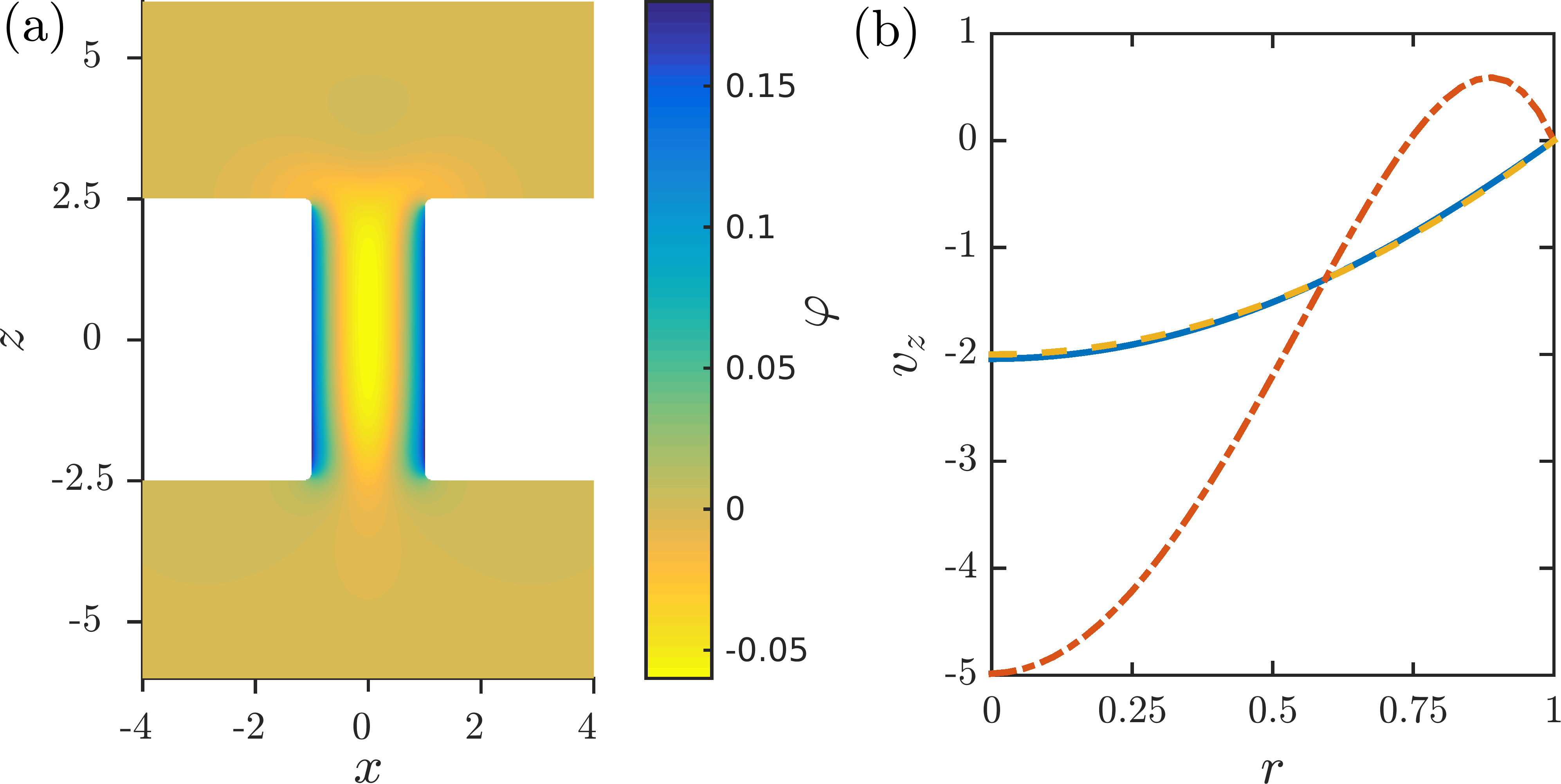}
\caption{(color online) (a) composition for case (ii) steady state. 
Parameters are the same as in \cref{fig2} but with $\up=5$ mm/s ($\pep=0.77$). 
(b) velocity profiles $v_z(r)$ at the pore center ($z=0)$ for a 
single-component fluid ($v_z=-2(1-r^2)$, dashed line) and for case (i) in 
\cref{fig2} and for case (ii) in this figure (dash-dot and solid lines, 
respectively)}
\label{fig5}
\end{figure}

Our results imply that solutes that 
interact favorably with the $\varphi<0$ 
phase will, for a wide range of applied pressures, be focused in the internal 
cylindrical region while being transported through the pore, and therefore, we 
speculate that the pore could exhibit anti-fouling behaviour 
\cite{hou2015}, depending on the solute-wall interaction strength. The circular 
cross 
section for which 
$\varphi<0$ is found at $r\approx 0.6\rp$ and this 
value is relatively insensitive to temperature. However, the affinity of 
solutes to the central 
region can be tuned externally via temperature.
As $T\rightarrow T_c$, $\vphi$ in the internal region becomes more negative.

Finally, when the pressure is even further increased,
steady state case (iii) is gradually reached. The adsorption in the pore 
further decreases 
until only a thin adsorbed layer near the wall remains. The excess 
material initially advected out of the pore is expelled into a reservoir 
where $\per$ is much smaller and is therefore dispersed slowly downstream. 
The result \cite{com_sup} is a steady state composition 
distribution in the reservoir that can extend over a huge distance of 
$O(\mu$m$)$. This regime will be investigated in more detail in a future 
publication. 

In conclusion, we predict that a tunable composition current can be pushed 
through a nanopore, and that the pore composition can be controlled reversibly 
by either temperature or pressure. The consequences of these phenomena on the 
transport of solutes through the pore is an intriguing possibility which we hope 
will motivate experimental work.

\begin{acknowledgments}
We acknowledge discussions with R. Evans and the anonymous referees for 
their useful comments. R.v.R acknowledges financial support of a Netherlands 
Organisation for 
Scientific Research (NWO) VICI grant funded by the Dutch Ministry
of Education, Culture and Science (OCW). S.S acknowledges funding from the 
European Union's Horizon 2020 research and innovation programme under the Marie 
Sk\l{}odowska-Curie grant agreement No. 656327. This work is part of the D-ITP 
consortium, a program of the Netherlands Organisation for Scientific
Research (NWO) funded by the Dutch Ministry of Education, Culture and Science 
(OCW). DNS in this work were performed using the COMSOL multiphysics 
software v5.1.
\end{acknowledgments}


\begin{thebibliography}{39}
\bibitem [{com({\natexlab{b}})}]{com_slip}%
  \BibitemOpen
  \href@noop {} {}\bibinfo {howpublished} {Our results remain qualitatively the
  same and quantitatively similar when a slip velocity up to 0.5$\up$ is
  prescribed at the pore wall.} ({\natexlab{b}})\BibitemShut {NoStop}%
\end{thebibliography}
\end{document}